\documentstyle[12pt]{article}

\begin{document}

\title{Sending quantum information with Gaussian states}
\author{A.S.Holevo \\
{\normalsize {\it Steklov Mathematical Institute, Russian Academy of
Sciences, Moscow.}}}
\date{}
\maketitle

\renewcommand{\theequation}{\arabic{section}. \arabic{equation}}

\section{Information characteristics of a quantum channel}

During the last couple of years an impressive progress has been achieved in
the theory of transmission of classical information through quantum
communication channels (see \cite{Hol} for a comprehensive survey). The
problem of sending quantum information is much less understood; we refer in
particular to the papers \cite{Lin}, \cite{Sch}, \cite{Cer}, initiating the
study of this problem, where the reader can find further references. In this
paper we make a contribution to this study by considering rather concrete
situation: sending Gaussian (quasifree) states through linear Bosonic
channels.

Consider quantum system in a Hilbert space ${\cal H}$, with a fixed density
operator $\rho .$ A{\it \ channel }is a transformation of quantum states as
presented by density operators, given by the relation

\[
T[\rho ]=\sum_{j}A_{j}\rho A_{j}^{*}, 
\]
where $A_{j}$ are bounded operators in ${\cal H}$ satisfying $%
\sum_{j}A_{j}^{*}A_{j}=I.$ Let us denote $H(\rho )=-{\rm Tr}\rho \log \,\rho 
$ the von Neumann entropy of a density operator $\rho .$ We call $\rho $ the
input state, and $T[\rho ]$ the output state of the channel. There are three
important entropy quantities related to the couple $(\rho ,T):$

1) The entropy of the input state $H(\rho );$

2) The entropy of the output state $H(T[\rho ]);$

3) The entropy exchange $H(\rho ,T).$

While the definition and the meaning of the first two entropies is obvious,
the third quantity is somewhat more sophisticated. To define it one
introduces the{\it \ reference system}, described by the Hilbert space $%
{\cal H}_{R},$ isomorphic to the Hilbert space ${\cal H}_{Q}=$ ${\cal H}$ of
the initial system. Then according to \cite{Lin}, \cite{Sch}$,$there exists 
{\it purification} of the state $\rho ,$ i.e. a unit vector $|\psi \rangle
\in {\cal H}_{Q}\otimes {\cal H}_{R}$ such that

\[
\rho ={\rm Tr}_{R}|\psi \rangle \langle \psi |. 
\]
The {\it entropy exchange} is then defined as

\[
H(\rho ,T)=H((T\otimes Id)[|\psi \rangle \langle \psi |]), 
\]
that is as the entropy of the output state of the dilated channel $(T\otimes
Id)$ applied to the input which is purification of the input state $\rho .$
One can then show that $H(\rho ,T)$ is equal to the entropy increase in the
channel environment $E$ provided the channel is represented by a unitary
interaction with the environment system being initially in a pure state \cite
{Lin}, \cite{Sch}$.$

From these three entropies one can construct three other quantities, bearing
some analogy with the classical mutual information. In general, if $\rho
_{12}$ is a density operator in a tensor product Hilbert space ${\cal H}%
_{1}\otimes {\cal H}_{2},$ and $\rho _{1},\rho _{2}$ are the partial states
of $\rho _{12},$resp., in ${\cal H}_{1},{\cal H}_{2},$then one can introduce
the quantity 
\[
C_{12}=H(\rho _{1})+H(\rho _{2})-H(\rho _{12}). 
\]
This quantity is nonnegative by subadditivity of quantum entropy, and to
certain extent reflects information correlation between the two systems \cite
{Lin}, although its operational meaning is still not completely clarified.
Then from the above three entropies one can construct three quantum
information quantities \cite{Lin}, \cite{Cer}:

1) {\it Quantum mutual information} 
\[
I=H(\rho )+H(T[\rho ])-H(\rho ,T), 
\]
reflecting quantum information transfer from the reference system $R$ to the
output of the initial system $Q^{\prime }.$ The important component of it is
the {\it coherent information} $H(T[\rho ])-H(\rho ,T)$, supremum of which
with respect to input states $\rho $ was conjectured as the quantum capacity
of the channel $T$ \cite{Sch}.

2) {\it Loss } 
\[
L=H(\rho )+H(\rho ,T)-H(T[\rho ]), 
\]
which reflects quantum information transfer from the reference system $R$ to
the output of the environment $E^{\prime }.$

3) {\it Noise} 
\[
N=H(T[\rho ])+H(\rho ,T)-H(\rho ),
\]
which reflects quantum information transfer from the output of the
environment $E^{\prime }$ to the output of the initial system $Q^{\prime }.$
In \cite{Cer} quantum Wenn's diagrams were introduced to visualize the
relations between the entropy and the information quantities. However, in
contrast to classical case, some areas in these diagrams representing
conditional entropies may have negative measure. We use another graphic
representation via the {\it information triangle}. In this representation
the entropies $H(\rho ),H(T[\rho ]),H(\rho ,T)$ are associated with the
sides of the triangle, and the information quantities $I,L,N$ are attached
to its vertices. The deficiency of this picture is that the representation
of the information quantities is only qualitative: roughly, the bigger is
the quantity - the bigger is distance from the corresponding vertex to the
opposite side of the triangle, and vice versa.

Although the entropy and information quantities described above were studied
in some detail from the general point of view, they are far from being
completely understood, and concrete examples in which they can be explicitly
evaluated are certainly welcome. In quantum statistics there is one large
class of states for which many explicit calculations are possible -- the so
called quasifree states of canonical commutation relations, in many respect
analogous to the classical Gaussian probability distributions. They are the
states of the maximal entropy among all states with fixed second moments,
for example, mean energy for a quadratic Hamiltonian. The aim of the present
paper is the study the behavior of the information triangle for Gaussian
input state and the most common attenuation/amplification channel.

\section{Quantum Gaussian states}

In this Section we repeat some results of \cite{Hol75}, \cite{Hol82}, \cite
{Sohma} and give a new variant of the expression for the entropy of a
general quantum Gaussian state. Let $q_{j},p_{j}$ be the canonical
observables satisfying the Heisenberg CCR 
\[
\lbrack q_{j},p_{k}]=i\delta _{jk}\hbar
I,\;\;[q_{j},q_{k}]=0,\;\;[p_{j},p_{k}]=0.
\]
Let us introduce the column vector 
\[
R=[q_{1},\dots ,q_{s};p_{1},\dots ,p_{s}]^{T}.
\]
We also introduce real column $2s$-vector $z=[x_{1},\dots ,x_{s};y_{1},\dots
,y_{s}]^{T}$, and the unitary operators in ${\cal H}$ 
\[
V(z)=\exp \,i\sum_{j=1}^{s}(x_{j}q_{j}+y_{j}p_{j})=\exp \,i\,R^{T}z.
\]
The operators $V(z)$ satisfy the Weyl-Segal CCR 
\begin{equation}
V(z)V(z^{\prime })=\exp [i/2\Delta (z,z^{\prime })]V(z+z^{\prime }),
\label{weyl}
\end{equation}
where 
\[
\Delta (z,z^{\prime })=\hbar \sum_{j=1}^{s}(x_{j}^{\prime
}y_{j}-x_{j}y_{j}^{\prime })
\]
is the canonical symplectic form. The Weyl -Segal CCR is the rigorous
counterpart of the Heisenberg CCR, involving only bounded operators. We
denote by

\begin{equation}
\Delta =\left[ 
\begin{array}{ccc}
\begin{array}{cccc}
0 &  &  &  \\ 
& 0 &  &  \\ 
&  & \ddots &  \\ 
&  &  & 0
\end{array}
& 
\begin{array}{c}
| \\ 
| \\ 
| \\ 
|
\end{array}
& 
\begin{array}{cccc}
\hbar &  &  & 0 \\ 
& \hbar &  &  \\ 
&  & \ddots &  \\ 
0 &  &  & \hbar
\end{array}
\\ 
\begin{array}{cccc}
- & - & - & -
\end{array}
& - & 
\begin{array}{cccc}
- & - & - & -
\end{array}
\\ 
\begin{array}{cccc}
-\hbar &  &  & 0 \\ 
& -\hbar &  &  \\ 
&  & \ddots &  \\ 
0 &  &  & -\hbar
\end{array}
& 
\begin{array}{c}
| \\ 
| \\ 
| \\ 
|
\end{array}
& 
\begin{array}{cccc}
0 &  &  &  \\ 
& 0 &  &  \\ 
&  & \ddots &  \\ 
&  &  & 0
\end{array}
\end{array}
\right]  \label{delta}
\end{equation}
the $(2s)\times (2s)$-skew-symmetric {\it commutation matrix} of components
of the vector $R$. Most of the results below are valid for the case where
the commutation matrix is arbitrary skew-symmetric matrix, not necessarily
of the canonical form (\ref{delta}).

{\bf \ }The density operator\ $\rho $ is called {\it Gaussian}, if its
quantum characteristic function has the form 
\[
{\rm Tr}\rho V(z)=\exp (i\,m^{T}z-\frac{1}{2}z^{T}\alpha z), 
\]
where $m$\ is column ($2s$)-vector and $\alpha $\ is real symmetric $%
(2s)\times (2s)$-matrix.

One can show that 
\[
m={\rm Tr}\rho R\;;\;\alpha -\frac{i}{2}\Delta ={\rm Tr}R\rho R^{T} 
\]
(cf{\it . }\cite{Hol75}, \cite{Hol82}). The {\it mean} $m$ can be arbitrary
vector; in what follows we will be interested in the case $m=0.$ The
necessary and sufficient condition on the {\it correlation matrix} $\alpha $
is the matrix uncertainty relation 
\begin{equation}
\alpha -\frac{i}{2}\Delta \geq 0.  \label{n-s condition}
\end{equation}
This condition is equivalent to its transpose $\alpha +\frac{i}{2}\Delta
\geq 0,$and to the following matrix generalization of the Heisenberg
uncertainty relation 
\begin{equation}
\Delta ^{-1}\alpha \Delta ^{-1}+\frac{1}{4}\alpha ^{-1}\geq 0,
\label{heisenberg}
\end{equation}
which is obtained by combining together (\ref{n-s condition}) and its
transpose. The state $\rho $ is pure if and only if the equality holds in
this equation, or 
\begin{equation}
(\Delta ^{-1}\alpha )^{2}=-\ \frac{1}{4}I.  \label{pure}
\end{equation}

Let us introduce the function

\[
g(x)=(x+1)\log (x+1)-x\log x,\qquad x>0. 
\]

We shall also use the matrix function abs$(\cdot )$, which is defined as
follows: for a diagonalizable matrix $M=T{\rm diag}(m_{j})T^{-1},$ we put $%
{\rm abs}M=T{\rm diag}(|m_{j}|)T^{-1}.$ In \cite{Sohma} it was shown that
the entropy of the Gaussian state is equal to 
\[
H(\rho )=\frac{1}{2}{\rm Sp}G(-(\Delta ^{-1}\alpha )^{2}),
\]
where 
\[
G(a^{2})=(a+\frac{1}{2})\log (a+\frac{1}{2})-(a-\frac{1}{2})\log (a-\frac{1}{%
2}),
\]
and Sp denotes trace of a matrix, as distinct from trace of operator. The
matrix $\ \Delta ^{-1}\alpha $ has purely imaginary eigenvalues $\pm ia_{j}$
and is diagonalizable. Since $G(a^{2})=g(|a|-\frac{1}{2}),$ we obtain
another expression

\begin{equation}
H(\rho )=\frac{1}{2}{\rm Sp}g({\rm abs}(\Delta ^{-1}\alpha )-\frac{I}{2}),
\label{abs}
\end{equation}
which will be used in the sequel.

\section{Purification of Gaussian states}

Let us denote ${\cal H}_{Q}={\cal H}_{1}$ the Hilbert space of irreducible
representation $z\rightarrow V_{1}(z)$ of the CCR (\ref{weyl}), ${\cal H}%
_{R}={\cal H}_{2}$ the Hilbert space of irreducible representation $%
z\rightarrow V_{2}(z)$ of the CCR

\[
V_{2}(z)V_{2}(z^{\prime })=\exp [-i/2\Delta (z,z^{\prime
})]V_{2}(z+z^{\prime }). 
\]
For example, $V_{2}(z)=\exp
\,i\sum_{j=1}^{s}(x_{j}p_{j}^{(2)}+y_{j}q_{j}^{(2)}),$ where $%
q_{j}^{(2)},p_{j}^{(2)}$ satisfy the Heisenberg CCR in ${\cal H}_{2}.$ In $%
{\cal H}_{1}\otimes {\cal H}_{2}$ the operators $V(z_{1},z_{2})=V_{1}(z_{1})%
\otimes V_{2}(z_{2})$ satisfy the CCR 
\[
V(z_{1},z_{2})V(z_{1}^{\prime },z_{2}^{\prime })=\exp [i/2\Delta
(z_{1},z_{2};z_{1}^{\prime },z_{2}^{\prime })]V(z_{1}+z^{\prime
}{}_{1},z_{2}+z^{\prime }{}_{2}), 
\]
where 
\[
\Delta (z_{1},z_{2};z_{1}^{\prime },z_{2}^{\prime })=\Delta
(z_{1},z_{1}^{\prime })-\Delta (z_{2},z_{2}^{\prime }). 
\]

Following \cite{Hol72} we introduce Gaussian state $\rho _{12}$ in ${\cal H}%
_{1}\otimes {\cal H}_{2}$ with the correlation matrix 
\[
\alpha _{12}=\left[ 
\begin{array}{ll}
\alpha & \Delta \sqrt{-(\Delta ^{-1}\alpha )^{2}-I/4} \\ 
-\Delta \sqrt{-(\Delta ^{-1}\alpha )^{2}-I/4} & \alpha
\end{array}
\right] , 
\]
that is 
\begin{eqnarray*}
&&{\rm Tr}\rho _{12}V(z_{1},z_{2}) \\
&=&\exp [-\frac{1}{2}(z_{1}^{T}\alpha z_{1}+z_{2}^{T}\alpha
z_{2}+z_{1}^{T}\Delta \sqrt{-(\Delta ^{-1}\alpha )^{2}-I/4}%
z_{2}-z_{2}^{T}\Delta \sqrt{-(\Delta ^{-1}\alpha )^{2}-I/4}z_{1})].
\end{eqnarray*}
Obviously, $\rho _{1}={\rm Tr}_{2}\rho _{12}.$

Let us show that $\rho _{12}$ is pure. With 
\[
\Delta _{12}=\left[ 
\begin{array}{ll}
\Delta & 0 \\ 
0 & -\Delta
\end{array}
\right] , 
\]
we have 
\[
\Delta _{12}^{-1}\alpha _{12}=\left[ 
\begin{array}{ll}
\Delta ^{-1}\alpha & \sqrt{-(\Delta ^{-1}\alpha )^{2}-I/4} \\ 
\sqrt{-(\Delta ^{-1}\alpha )^{2}-I/4} & -\Delta ^{-1}\alpha
\end{array}
\right] , 
\]
and it is easy to check that $(\Delta _{12}^{-1}\alpha _{12})^{2}=-I/4.$ By
the criterium (\ref{pure}) $\rho _{12}$ is pure.

We shall be interested in the particular subclass of Gaussian states most
familiar in quantum optics, which we call gauge-invariant. These are the
states having the P-representation 
\begin{equation}
\rho =\pi ^{-s}|\det N|^{-1}\int \exp (-\zeta ^{\dagger }N^{-1}\zeta )|\zeta
\rangle \langle \zeta |d^{2s}\zeta  \label{p-repr}
\end{equation}
(see e.g. \cite{Hel}, Sec. V, 5. II). Here $\zeta \in {\bf C}^{s}$, $|\zeta
\rangle $are the coherent vectors in ${\cal H}$, $a|\zeta \rangle =\zeta
|\zeta \rangle $, $N$ is positive Hermitian matrix such that 
\[
N={\rm Tr}a\,\rho \,a^{\dagger } 
\]
(we use here vector notations, where $a=[a_{1},\dots ,a_{s}]^{T}$ is a
column vector and $a^{\dagger }=[a_{1}^{\dagger },\dots ,a_{s}^{\dagger }]$
is a row vector) and $a_{j}=\frac{1}{2\hbar }(q_{j}+ip_{j}).$As shown in 
\cite{Sohma}, the correlation matrix of such states is 
\[
\alpha =\hbar \left[ 
\begin{array}{ll}
{\rm Re}N+I/2 & -{\rm Im}N \\ 
{\rm Im}N & {\rm Re}N+I/2
\end{array}
\right] , 
\]

The real $2s\times 2s-$ matrices of such form can be rewritten as complex $%
s\times s-$ matrices, by using the correspondence 
\[
\left[ 
\begin{array}{ll}
A & -B \\ 
B & A
\end{array}
\right] \leftrightarrow A+iB, 
\]
which is in fact algebraic isomorphism, provided $A^{T}=A,B^{T}=-B.$
Apparently, 
\[
\frac{1}{2}{\rm Sp}\left[ 
\begin{array}{ll}
A & -B \\ 
B & A
\end{array}
\right] ={\rm Sp} (A+iB). 
\]

By using this correspondence, we have 
\[
\alpha \leftrightarrow \hbar (N+I/2),\qquad \Delta \leftrightarrow -i\hbar
I, 
\]
and 
\[
\Delta ^{-1}\alpha \leftrightarrow i(N+I/2). 
\]
In particular, the formula (\ref{abs}) becomes 
\[
H(\rho )={\rm Sp}g(N), 
\]
which is well known (see e.g. \cite{Sohma}) and confirms (\ref{abs}).

For future use we also need the correspondence 
\begin{equation}
\Delta _{12}^{-1}\alpha _{12}\leftrightarrow \left[ 
\begin{array}{ll}
i(N+I/2) & \sqrt{N^{2}+N} \\ 
\sqrt{N^{2}+N} & -i(N+I/2)
\end{array}
\right] .  \label{d-1a}
\end{equation}
For the case of one degree of freedom we shall be interested in the
following Section, $N$ is just nonnegative number and $\rho $ is {\it %
elementary} Gaussian state with the characteristic function 
\begin{equation}
\exp \left[ -\frac{\hbar }{2}\left( N+\frac{1}{2}\right) |z|^{2}\right] ,
\label{one mode charc. fct. }
\end{equation}
where we put $|z|^{2}=(x^{2}+y^{2}).$

\section{Attenuation/amplification channel}

Let us consider CCR with one degree of freedom described by one mode
annihilation operator $a=\frac{1}{2\hbar }(q+ip),$ and let $a_{0}$ be
another mode in the Hilbert space ${\cal H}_{0}={\cal H}_{E}$ describing
environment. Let the environment be initially in the vacuum state, which is
described by the characteristic function (\ref{one mode charc. fct. }) with $%
N=0$ i.e. $\exp [-\frac{\hbar }{4}|z|^{2}],$

The linear attenuator with coefficient $k<1$ is described by the
transformation 
\[
a\rightarrow ka+\sqrt{1-k^{2}}a_{0} 
\]
in the Heisenberg picture. Similarly, the linear amplifier with coefficient $%
k>1$ is described by the transformation 
\[
a\rightarrow ka+\sqrt{k^{2}-1}a_{0}. 
\]
It follows that the corresponding transformations $T_{k}[\rho ]$ of states
in the Schroedinger picture have, correspondingly, the characteristic
functions 
\begin{equation}
{\rm Tr}T_{k}[\rho ]V(z)={\rm Tr}\rho V(kz)\exp [-\frac{\hbar }{4}%
(1-k^{2})|z|^{2}],\qquad k<1,  \label{atten}
\end{equation}
\begin{equation}
{\rm Tr}T_{k}[\rho ]V(z)={\rm Tr}\rho V(kz)\exp [-\frac{\hbar }{4}%
(k^{2}-1)|z|^{2}],\qquad k>1,  \label{amplif}
\end{equation}
see \cite{Hol72a}. Let the input state $\rho $ of the system have the
characteristic function (\ref{one mode charc. fct. }), i.e. $\exp [-\frac{%
\hbar }{4}(N+\frac{1}{2})|z|^{2}].$ The entropy of $\rho $ is 
\[
H(\rho )=g(N). 
\]

From (\ref{atten}), (\ref{amplif}) we find that the output state $T_{k}[\rho
]$ is again elementary Gaussian with $N$ replaced by

\[
\widetilde{N}=k^{2}N,\qquad k<1;\qquad \widetilde{N}=k^{2}N+(k^{2}-1),\qquad
k>1. 
\]
Thus 
\[
H(T_{k}[\rho ])=g(k^{2}N),\qquad k<1;\qquad H(T_{k}[\rho
])=g(k^{2}N+(k^{2}-1)),\qquad k>1. 
\]

Now we calculate the entropy exchange $H(\rho ,T_{k}).$ The (pure) input
state $\rho _{12}$ of the extended system ${\cal H}_{1}\otimes {\cal H}_{2}$
is characterized by the $2\times 2-$matrix (\ref{d-1a}). The action of the
extended channel $(T\otimes Id)$ transforms this matrix into 
\[
\Delta _{12}^{-1}\widetilde{\alpha }_{12}\leftrightarrow \left[ 
\begin{array}{ll}
i(\widetilde{N}+\frac{1}{2}) & k\sqrt{N^{2}+N} \\ 
k\sqrt{N^{2}+N} & -i(N+\frac{1}{2})
\end{array}
\right] .
\]
From formula (\ref{abs}) we deduce $H(\rho ,T_{k})=g(|\lambda _{1}|-\frac{1}{%
2})+g(|\lambda _{2}|-\frac{1}{2}),$ where $\lambda _{1},\lambda _{2}$ are
the eigenvalues of the complex matrix in the right-hand side. The
eigenvalues are: $\lambda _{1}=\frac{i}{2},$%
\[
\lambda _{2}=-i[(1-k^{2})N+\frac{1}{2}],\qquad k<1;\qquad i[(k^{2}-1)(N+1)+%
\frac{1}{2}],\qquad k>1.
\]
Therefore we obtain 
\[
H(\rho ,T_{k})=g((1-k^{2})N)\qquad k<1;\qquad g((k^{2}-1)(N+1)),\qquad k>1.
\]

The behavior of the entropies $H(T_{k}[\rho ]),H(\rho ,T_{k})$ as functions
of $k$ is clear from Fig.1 . In particular, for all $N$ the coherent
information $H(T_{k}[\rho ])-H(\rho ,T_{k})$ turns out to be positive for $%
k>1/\sqrt{2}$ and negative otherwise. It tends to $-H(\rho )$ for $%
k\rightarrow 0,$ is equal to $H(\rho )$ for $k=1,$and quickly tends to zero
as $k\rightarrow \infty $ (Fig.2; on both plots $N=1$). The behavior of the
information triangle shows that loss dominates for $k\rightarrow 0,$ mutual
information for $k\div 1,$ while the noise -- as $k\rightarrow \infty .$
This agrees with what one should expect on physical grounds from quantities
presenting quantum mutual information, loss and noise and gives further
support for their use in quantum information theory. However, negativity of
the coherent information for $k<1/\sqrt{2}$ looks somewhat mysterious and
waits for a physical explanation.

Acknowledgments. This work was initiated when the author was visiting
Physical Department of the University of Milan under the contract with
Italian Ministry of Foreign Affairs provided by A. Volta Center for
Scientific Culture. The author is grateful to Prof. L.\ Lanz for his
hospitality and stimulating discussions. Part of this work was completed
during the 1998 Elsag-Bailey -- I.S.I. Foundation research meeting on
quantum computation.


\begin{thebibliography}{99}
\bibitem{Cer}  C. Adami and N. J. Cerf,\newblock ``Capacity of noisy quantum
channels'', \newblock {\it Phys. Rev. A}, vol. {\bf A56}, pp. 3470-3485,
1972. 

\bibitem{Sch}  H. Barnum, M. A. Nielsen, B. Schumacher, \newblock %
``Information transmission through noisy quantum channels'',\newblock  LANL
Report no. quant-ph/9702049, Feb. 1997.

\bibitem{Hel}  C. ~W. ~Helstrom, \newblock {\it Quantum detection and
estimation theory}, chapter 5, Academic press, 1976. 

\bibitem{Hol72}  A. ~S. ~Holevo,\newblock ``Generalized free states of the C$%
^{*}$-algebra of the CCR. '', \newblock {\it Theor. Math. Phys.}, vol. {\bf 6%
}, no.1, pp. 3-20, 1971. 

\bibitem{Hol72a}  A. ~S. ~Holevo,\newblock ``Towards the mathematical theory
of quantum communication channels'', \newblock {\it Problems of Information
Transm.}, vol. {\bf 8}, no.1, pp. 63-71, 1972. 

\bibitem{Hol75}  A. ~S. ~Holevo, \newblock ``Some statistical problems for
quantum Gaussian states'', \newblock {\it IEEE Transactions on Information
Theory}, vol. {\bf IT-21}, no.5, pp. 533-543, 1975. 

\bibitem{Hol82}  A. ~S. ~Holevo, \newblock {\it Probabilistic and
statistical aspects of quantum theory}, chapter 5, North-Holland, 1982. 

\bibitem{Hol}  A. ~S. ~Holevo, \newblock ``Coding theorems for Quantum
Channels'', \newblock {\it Tamagawa University Research Review}, No.4, 1998. 

\bibitem{Sohma}  A. ~S. ~Holevo, M. ~Sohma and O. ~Hirota, \newblock ``The
capacity of quantum Gaussian channels '', \newblock  Preprint 1998.

\bibitem{Lin}  G. Lindblad, \newblock ``Quantum entropy and quantum
measurements'', \newblock  {\it Lect. Notes Phys.}, vol. {\bf 378}, Quantum
Aspects of Optical Communication, Ed. by C. Benjaballah, O. Hirota, S.
Reynaud, pp.71-80, 1991.
\end{thebibliography}
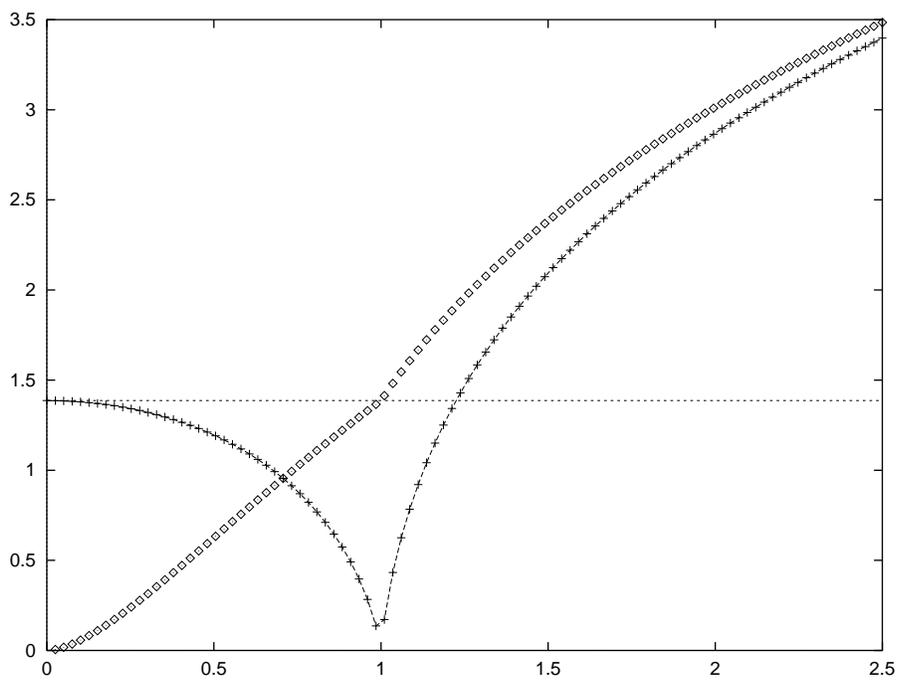
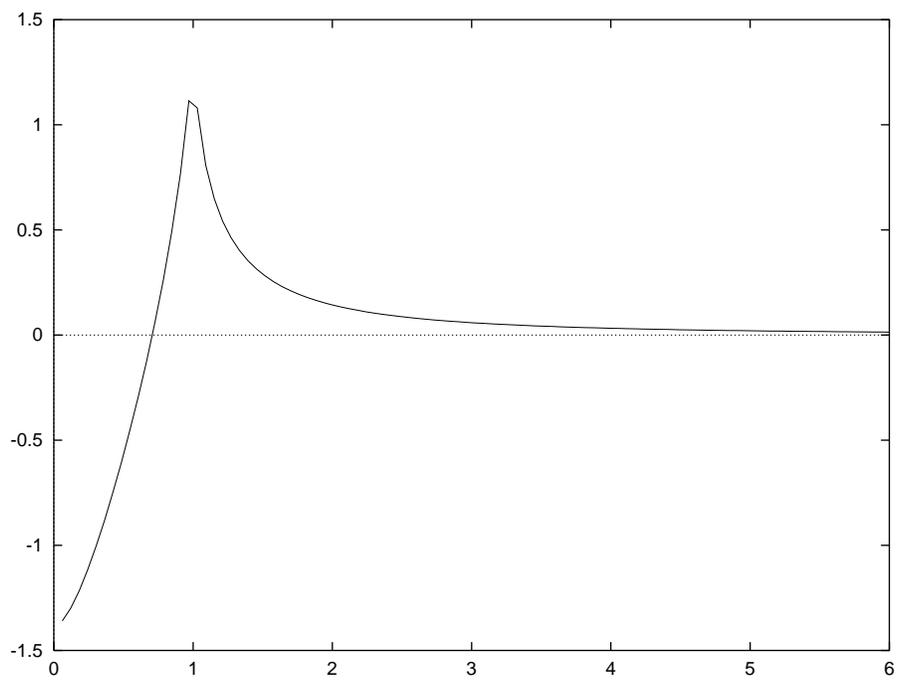
\end{document}